\documentstyle[multicol,aps,epsf]{revtex}
\tightenlines
\begin{document}
\draft \title{Domain motion in confined liquid crystals} 
\author{Colin Denniston$^1$, G\'eza T\'oth$^2$ and J.M. Yeomans$^2$}
\address{$^1$ Dept. of Physics and Astronomy, The Johns Hopkins
  University, Baltimore, MD 21218.}
\address{$^2$ Dept. of Physics, Theoretical Physics,
University of Oxford, 1 Keble Road, Oxford OX1 3NP}

\date{\today} \maketitle

\begin{abstract}
We extend a lattice Boltzmann algorithm of liquid crystal
hydrodynamics to include an applied electric field. The approach
solves the equations of motion written in terms of a tensor order
parameter. Back-flow effects and the hydrodynamics of topological
defects are included.  We investigate some of the dynamics relevant to
liquid crystal devices; in particular defect-mediated motion of domain
walls relevant to the nucleation of states useful in pi-cells. 
An anisotropy in the domain wall velocity is seen because defects
of different topology couple differently to the flow field. 
\end{abstract}

\section{Introduction}

The coupling of the optic and electric response of liquid crystals
has lead to their wide application in display devices in recent years.
Liquid crystalline materials are often made up of long, thin, rod-like 
molecules\cite{GP93}.  This anisotropy is what leads to their useful
optic properties.  The molecular geometry and interactions can lead to
a wide range of equilibrium phases.  In this paper we are concerned
with the nematic phase where the molecules tend to align along a
preferred direction, referred to as the director, giving long-range
orientational order.  The propensity to order, as well as the direction
along which the system orders are conveniently described by a tensor
order parameter ${\bf Q}$ \cite{GP93}.

An important feature of the nematic phase is the possibility of
topological defects.  These are director configurations that cannot
relax to the nematic ground state by a continuous rotation of the
molecules.  Examples of such defects that we will consider are shown
in Fig.~\ref{fig_twodefects}(b).
In addition, although widely recognized as an important factor
in device performance, the flow of the liquid crystal in response to
changes in the orientation of the molecules is both difficult to
measure experimentally and to incorporate into a simulation of the device.

Liquid crystal hydrodynamics are typically described by the
Ericksen-Leslie-Parodi equations of motion \cite{GP93}. However these
are restricted to an order parameter field of constant magnitude
and do not include the hydrodynamics of topological defects. 
Here we consider the Beris-Edwards formulation of liquid crystal
hydrodynamics\cite{BE94_BE90}. This allows for variations in the
magnitude of the nematic order parameter and correctly models both
defect dynamics and the coupling between the velocity field and 
the motion of the order parameter.

The Beris-Edwards equations are complex and previous numerical work to
explore different flow regimes is very limited. In Ref.\cite{RT98}
these equations were used but the problem was simplified by imposing a
velocity field and ignoring back-flow (back-flow refers to the effect of
the order parameter dynamics on the velocity field).  In \cite{F98} an
Euler solution of a somewhat simpler version of the Beris-Edwards
model was used to study the effect of hydrodynamics on phase ordering
in liquid crystals.  However it has recently been shown that the
Beris-Edwards equations can be solved successfully using a lattice
Boltzmann algorithm \cite{DO00}.

In this paper we will extend the lattice Boltzmann approach described
in \cite{DO00} to include the electric field terms necessary for a
description of electro-optic device physics.  The model is also
generalized to include the three elastic constant formula for
the deformation free energy of the liquid crystal.  The equations of
motion are presented in Section II and an outline of the numerical
algorithm in Section III.

We then apply the algorithm to some examples of
interest in device physics \cite{S00}.  The nematic liquid crystal is confined
between two plates a few $\mu m$ apart.  The director configuration
on the plates is fixed. When an electric field is switched on, the
molecules align parallel or perpendicular to the field depending on
the sign of the anisotropy of the dielectric constant. 
After switching off the field, long-range elastic interactions ensure 
that the molecules reorient themselves in the direction preferred by
the surfaces.  The device can be used as a display because different
liquid crystal orientations have different optical properties.

Switching between different director orientations can take
place homogeneously throughout the bulk of the device or it may
proceed through domain growth involving the motion of 
domain walls, together with associated defects \cite {AT00}. 
The operational state of a device such as a pi-cell may be
topologically distinct from its state at zero voltage.  Before the
device can be used, the operational state must be nucleated and must
grow to fill the display. It is important to examine how this 
depends on different parameters of the liquid crystal material. In
particular  liquid crystals are complex fluids and the speed of domain
motion will depend on their hydrodynamic properties.  

In section IV we examine the global stability of two topologically
distinct states as a function of the applied voltage and the tilt
angle at the plates.  In Section V we discuss the relevance of defect
motion to domain growth between the two states and show that
hydrodynamic coupling significantly changes the defect velocities.

\section{The hydrodynamic equations of motion}
\label{2.0}

We summarize the formulation of liquid crystal hydrodynamics
described by Beris and Edwards\cite{BE94_BE90} and extended to include
electric fields.  The continuum equations of motion are written
in terms of a tensor order parameter ${\bf Q}$ which is related to the
direction of individual molecules ${\vec{m}}$ by 
$Q_{\alpha\beta}= \langle m_\alpha m_\beta -
{1\over 3} \delta_{\alpha\beta}\rangle$ where the angular brackets
denote a coarse-grained average. (Greek indices will be
used to represent Cartesian components of vectors and tensors 
and the usual summation over
repeated indices will be assumed.) ${\bf Q}$ is a traceless symmetric
tensor. Its largest eigenvalue, $\frac {2} {3} q$, $0<q<1$, 
describes the magnitude of the order.

We first write down a Landau
free energy which describes the equilibrium properties of the liquid 
crystal. This appears in the 
equation of motion of the order parameter, which includes a
Cahn-Hilliard-like term through which the system evolves towards 
thermodynamic equilibrium. 

{\bf Free energy:}
The equilibrium properties of a nematic liquid crystal can be
described by the Landau-de Gennes free energy\cite{GP93,P91}
\begin{equation}
{\cal F}=\int_{V} dV \left\{ f_{bulk}+f_{elastic}+f_{field}\right\} + 
         \int_{S} dS \left\{ f_{surf}\right\}.
\label{free}
\end{equation}
$f_{bulk}$ describes the bulk free energy\
\begin{equation}
f_{bulk}=\frac {A_0}{2} (1 - \frac {\gamma} {3}) Q_{\alpha \beta}^2 - 
          \frac {A_0 \gamma}{3} Q_{\alpha \beta}Q_{\beta \gamma}Q_{\gamma \alpha} + 
          \frac {A_0 \gamma}{4} (Q_{\alpha \beta}^2)^2.
\end{equation}
For $\gamma=2.7$ there is a first-order transition from the isotropic to the nematic phase. 
$f_{elastic}$ is the analogue of the Frank elastic free energy density
\begin{equation}
f_{elastic}=\frac{L_1}{2} (\partial_\alpha Q_{\beta \gamma})^2+
\frac{L_2}{2} (\partial_\alpha Q_{\alpha \gamma})(\partial_\beta Q_{\beta \gamma})+
\frac{L_3}{2} Q_{\alpha \beta}(\partial_\alpha Q_{\gamma \epsilon})
(\partial_\beta Q_{\gamma \epsilon}).
\label{fFrank}
\end{equation}

The dielectric constant is usually measured along and perpendicular to
the nematic axis, $\vec{n}$, and is usually 
assumed to give a relation between the
electric displacement ${\bf D}$ and field ${\bf E}$ of the form
\begin{equation}
{\bf D}=\epsilon_\perp {\bf E}+(\epsilon_\parallel-\epsilon_\perp)({\vec
  n}\cdot{\bf E}) {\vec n}.
\end{equation} 
For a uniaxial nematic, this is entirely equivalent to assuming that
\begin{equation}
\epsilon_{\alpha\beta}=\frac{2}{3}\epsilon_a
Q_{\alpha\beta}+\epsilon_m \delta_{\alpha\beta}
\end{equation}
where
\begin{eqnarray}
\epsilon_a &=& \frac{3}{2 q} (\epsilon_\parallel-\epsilon_\perp),\\
\epsilon_m &=& \frac{2}{3}\epsilon_\perp+ \frac{1}{3}\epsilon_\parallel
\end{eqnarray}
The electric contribution to the thermodynamic potential $f_{field}$  is
\begin{equation}
f_{field}= -\frac{1}{4\pi}\int {\bf D}\cdot d{\bf E}=-\frac{1}{8\pi}\epsilon_m
E^2-\frac{1}{12\pi}\epsilon_a E_\alpha E_\beta Q_{\alpha\beta}.
\end{equation}

At the surfaces of the device we assume a pinning potential
\begin{equation}
f_{surf}= \frac {1} {2} \alpha_S (Q_{\alpha \beta}-
                Q_{\alpha \beta}^0)^2.
\label{free_surface}
\end{equation}
This corresponds to a director at the surface preferring to lie along
the direction of the eigenvector of ${\bf Q}^0$ corresponding to the
largest eigenvalue $2/3 q$.

{\bf Equation of motion of the nematic order parameter:}
The equation of motion for the nematic order parameter is\cite{BE94_BE90}
\begin{equation}
(\partial_t+{\vec u}\cdot{\bf \nabla}){\bf Q}-{\bf S}({\bf W},{\bf
  Q})= \Gamma {\bf H}
\label{Qevolution}
\end{equation}
where $\Gamma$ is a collective rotational diffusion constant.
The first term on the left-hand side of equation (\ref{Qevolution})
is the material derivative describing the usual time dependence of a
quantity advected by a fluid with velocity ${\vec u}$. This is
generalized by a second term 
\begin{eqnarray}
{\bf S}({\bf W},{\bf Q})
&=&(\xi{\bf \Lambda}+{\bf \Omega})({\bf Q}+{\bf I}/3)+({\bf Q}+
{\bf I}/3)(\xi{\bf \Lambda}-{\bf \Omega})\nonumber\\
& & -2\xi({\bf Q}+{\bf I}/3){\mbox{Tr}}({\bf Q}{\bf W})
\end{eqnarray}
where ${\bf \Lambda}=({\bf W}+{\bf W}^T)/2$ and
${\bf \Omega}=({\bf W}-{\bf W}^T)/2$
are the symmetric part and the anti-symmetric part respectively of the
velocity gradient tensor $W_{\alpha\beta}=\partial_\beta u_\alpha$.
${\bf S}({\bf W},{\bf Q})$  appears in the equation of motion because
the order parameter distribution can be both rotated and stretched by
flow gradients. This is the consequence of the rod-like geometry
of the liquid crystal molecules.
$\xi$ is a constant which will depend on the molecular
details of a given liquid crystal.

The term on the right-hand side of equation (\ref{Qevolution})
describes the relaxation of the order parameter towards the minimum of
the free energy. The molecular field ${\bf H}$ which provides the driving
motion is related to the derivative of the free energy by
\begin{eqnarray}
{\bf H}&=& -{\delta {\cal F} \over \delta {\bf Q}}+({\bf
    I}/3) Tr{\delta {\cal F} \over \delta {\bf Q}}\nonumber\\ &=&
    {\bf H_{bulk}}+{\bf H_{elastic}}+{\bf H_{field}}+{\bf H_{surf}},
\label{H(Q)}
\end{eqnarray}  
where 
\begin{eqnarray}
{\bf H_{bulk}}  &=& - A_0 (1 - \frac {\gamma} {3})
 {\bf Q}+ A_0 \gamma \left({\bf Q^2}-
({\bf I}/3)Tr{\bf Q^2}\right)- 
A_0 \gamma {\bf Q}Tr{\bf Q^2},\\
H_{elastic,\alpha \beta} &=& L_1 ({\partial_\gamma}^2 Q_{\alpha \beta}) \nonumber\\
& & + L_2 \left \{ \frac {1} {2}
(\partial_{\alpha} \partial_{\gamma} Q_{\gamma \beta}  +
 \partial_{\beta}  \partial_{\gamma} Q_{\gamma \alpha}) -
 \frac {1} {3} \delta_{\alpha \beta} 
\partial_{\gamma} \partial_{\epsilon} Q_{\gamma \epsilon}\right \}  \nonumber\\
& & + \frac {1} {2} L_3 \biggl \{ \partial_\gamma 
( Q_{\gamma \epsilon} \partial_{\epsilon} 
Q_{\alpha \beta})
-(\partial_\alpha Q_{\gamma \epsilon})
(\partial_\beta Q_{\gamma \epsilon})
+ \frac {1} {3} \delta_{\alpha \beta} (\partial_\eta Q_{\gamma \epsilon})^2
\label{H_elastic}
\biggr \},\\
H_{field,\alpha \beta} &=& \frac {\epsilon_a} {12 \pi} 
( E_{\alpha } E_{\beta}  - \frac {\delta_{\alpha \beta}} {3} {E_{\gamma}}^2),\\
{\bf H_{surf}} &=& -\alpha_S({\bf Q}-{\bf Q}^{0}),
\end{eqnarray}
and $\delta_{\alpha \beta}$ is the Kronecker delta. 
For (\ref{H_elastic}) 
$\partial_z=0$ is assumed, and the symmetry and tracelessness of ${\bf
  Q}$ is also exploited for simplification.  

When computing the functional derivative in (\ref{H(Q)}) surface terms
arise from the integration by parts.  For reference, these are
\begin{equation}
\delta{\cal F}_d^{surf} = \int_{\partial V} ds\, \delta Q_{\alpha\beta} \left\{
L_1 \sigma_\gamma (\partial_\gamma Q_{\alpha\beta})
+\frac{1}{2}L_2 [\sigma_\alpha (\partial_\gamma Q_{\gamma\beta})
+ \sigma_\beta (\partial_\gamma Q_{\gamma\alpha})]
+L_3 \sigma_\gamma Q_{\gamma\epsilon}(\partial_\epsilon Q_{\alpha\beta})
\right\},
\end{equation}
where ${\bf \sigma}$ is the surface (unit) normal.

{\bf Continuity and Navier-Stokes equations:}
The fluid momentum obeys the continuity
\begin{equation}
\partial_t \rho + \partial_{\alpha} \rho u_{\alpha} =0,
\label{continuity}
\end{equation}
and the Navier-Stokes equation 
\begin{equation}
\rho\partial_t u_\alpha+\rho u_\beta \partial_\beta
u_\alpha=\partial_\beta \tau_{\alpha\beta}+\partial_\beta
\sigma_{\alpha\beta}+{\rho \tau_f \over
3}\partial_\beta((\delta_{\alpha \beta}-3\partial_\rho
P_{0})\partial_\gamma u_\gamma+\partial_\alpha
u_\beta+\partial_\beta u_\alpha),
\label{NS}
\end{equation}
where $\rho$ is the fluid density and $\tau_f$ is related to the viscosity.
The form of this equation is not dissimilar to that for a simple
fluid. However the details of the stress tensor reflect the additional
complications of liquid crystal hydrodynamics. 
There is a symmetric contribution
\begin{eqnarray}
\sigma_{\alpha\beta} &=&-P_0 \delta_{\alpha \beta}
-\xi H_{\alpha\gamma}(Q_{\gamma\beta}+{1\over
  3}\delta_{\gamma\beta})-\xi (Q_{\alpha\gamma}+{1\over
  3}\delta_{\alpha\gamma})H_{\gamma\beta}\nonumber\\
& & \quad +2\xi
(Q_{\alpha\beta}+{1\over 3}\delta_{\alpha\beta})Q_{\gamma\epsilon}
H_{\gamma\epsilon}-\partial_\beta Q_{\gamma\nu} {\delta
{\cal F}\over \delta\partial_\alpha Q_{\gamma\nu}} +\sigma_{M,\alpha\beta}
\label{BEstress}
\end{eqnarray}
and an antisymmetric contribution
\begin{equation}
 \tau_{\alpha \beta} = Q_{\alpha \gamma} H_{\gamma \beta} -H_{\alpha
 \gamma}Q_{\gamma \beta}.
\label{as}
\end{equation}
The background pressure $P_0$ is taken simply to be
\begin{equation}
P_0 = \rho T,
\end{equation}
and is found to be a constant in our simulations to a very good approximation.
The Maxwell stress tensor $\sigma_{M,\alpha\beta}$ 
\cite{LandauLifshitz} describes the stress due to the external electric field
\begin{equation}
\sigma_{M,\alpha\beta} = \frac {1} {8 \pi} ( D_\alpha E_\beta +  E_\alpha D_\beta  
                         - D_\gamma E_\gamma \delta_{\alpha \beta}).
\end{equation}

\section{A lattice Boltzmann algorithm for liquid crystal
hydrodynamics}

We now describe a lattice Boltzmann algorithm which solves the
hydrodynamic equations of motion of a liquid crystal 
(\ref{Qevolution}), (\ref{continuity}), and (\ref{NS}). 
Lattice Boltzmann algorithms are defined in
terms of a set of continuous variables, usefully termed partial
distribution functions, which move on a lattice in discrete space and
time \cite{CD98}.

The simplest lattice Boltzmann algorithm, which solves the
Navier-Stokes equations of a simple fluid, is defined in terms of a
single set of partial distribution functions which sum on each site to
give the density. For liquid crystal hydrodynamics this must be
supplemented by a second set, which are tensor variables, and which
are related to the tensor order parameter ${\bf Q}$ \cite{DO00}.

We define two distribution functions, the scalars $f_i (\vec{x})$ and
the symmetric traceless tensors ${\bf G}_i (\vec{x})$ on each lattice
site $\vec{x}$. Each $f_i$, ${\bf G}_i$ is associated with a lattice
vector ${\vec e}_i$. We choose a nine-velocity model on a square
lattice with velocity vectors ${\vec e}_i=(\pm 1,0),(0,\pm 1), (\pm 1, \pm
1), (0,0)$. Physical variables are defined as moments of the
distribution function
\begin{equation}
\rho=\sum_i f_i, \qquad \rho u_\alpha = \sum_i f_i  e_{i\alpha},
\qquad {\bf Q} = \sum_i {\bf G}_i.
\label{eq1}
\end{equation} 

The distribution functions evolve in a time step $\Delta t$ according to
\begin{eqnarray}
&&f_i({\vec x}+{\vec e}_i \Delta t,t+\Delta t)-f_i({\vec x},t)=
\frac{\Delta t}{2} \left[{\cal C}_{fi}({\vec x},t,\left\{f_i
\right\})+ {\cal C}_{fi}({\vec x}+{\vec e}_i \Delta
t,t+\Delta
t,\left\{f_i^*\right\})\right],
\label{eq2}\\
&&{\bf G}_i({\vec x}+{\vec e}_i \Delta t,t+\Delta t)-{\bf G}_i({\vec
x},t)= \nonumber\\
&& \qquad\qquad \qquad\qquad \qquad\qquad\frac{\Delta t}{2}\left[ {\cal C}_{{\bf G}i}({\vec
x},t,\left\{{\bf G}_i \right\})+
                {\cal C}_{{\bf G}i}({\vec x}+{\vec e}_i \Delta
                t,t+\Delta t,\left\{{\bf G}_i^*\right\})\right].
\label{eq3}
\end{eqnarray}
This represents free streaming with velocity ${\vec e}_i$ and a
collision step which allows the distribution to relax towards
equilibrium. 
$f_i^*$ and ${\bf G}_i^*$ are first order approximations to 
$f_i({\vec x}+{\vec e}_i \Delta t,t+\Delta t)$ and ${\bf G}_i({\vec
  x}+{\vec e}_i \Delta t,t+\Delta t)$
respectively. 
Discretizing in this way, which is similar to a predictor-corrector 
scheme, means that lattice viscosity terms are eliminated.
In a lattice Boltzmann algorithm for a simple fluid these
terms can be added to the physical viscosity term.
Here this is not the case: the lattice viscosity gives an additional,
spurious term in the equations of motion.  
An additional advantage of the predictor-corrector approach is
that the stability of the lattice Boltzmann algorithm is
improved.
The collision operators are taken to have the form of a single
relaxation time Boltzmann equation\cite{CD98}, together with a forcing term
\begin{eqnarray}
{\cal C}_{fi}({\vec x},t,\left\{f_i \right\})&=&
-\frac{1}{\tau_f}(f_i({\vec x},t)-f_i^{eq}({\vec x},t,\left\{f_i
\right\}))
+p_i({\vec x},t,\left\{f_i \right\}),
\label{eq4}\\
{\cal C}_{{\bf G}i}({\vec x},t,\left\{{\bf G}_i
\right\})&=&-\frac{1}{\tau_{g}}({\bf G}_i({\vec x},t)-{\bf
G}_i^{eq}({\vec x},t,\left\{{\bf G}_i \right\}))
+{\bf M}_i({\vec x},t,\left\{{\bf G}_i \right\}).
\label{eq5}
\end{eqnarray}

The form of the equations of motion and thermodynamic equilibrium
follow from the choice of the moments of the equilibrium distributions
$f^{eq}_i$ and ${\bf G}^{eq}_i$ and the driving terms $p_i$ and
${\bf M}_i$. $f_i^{eq}$ is constrained by
\begin{equation}
\sum_i f_i^{eq} = \rho,\qquad \sum_i f_i^{eq} e_{i \alpha} = \rho
u_{\alpha}, \qquad
\sum_i f_i^{eq} e_{i\alpha}e_{i\beta} = -\sigma_{\alpha\beta}+\rho
u_\alpha u_\beta
\label{eq6} 
\end{equation}
where the zeroth and first moments 
are chosen to impose conservation of
mass and momentum. The second moment of $f^{eq}$ controls the symmetric
part of the stress tensor, whereas the moments of $p_i$
\begin{equation}
\sum_i p_i = 0, \quad \sum_i p_i e_{i\alpha} = \partial_\beta
\tau_{\alpha\beta},\quad \sum_i p_i
e_{i\alpha}e_{i\beta} = 0
\label{eq7}
\end{equation}
impose the antisymmetric part of the stress tensor.
For the equilibrium of the order parameter distribution we choose
\begin{equation}
\sum_i {\bf G}_i^{eq} = {\bf Q},\qquad \sum_i
{\bf G}_i^{eq} {e_{i\alpha}} = {\bf Q}{u_{\alpha}},
\qquad \sum_i {\bf G}_i^{eq}
e_{i\alpha}e_{i\beta} = {\bf Q} u_\alpha u_\beta .
\label{eq8}
\end{equation}
This ensures that the order parameter
is convected with the flow. Finally the evolution of the
order parameter is most conveniently modeled by choosing
\begin{equation}
\sum_i {\bf M}_i = \Gamma {\bf H}({\bf Q})
+{\bf S}({\bf W},{\bf Q}) \equiv {\bf \hat{H}}, \qquad
\qquad \sum_i {\bf M}_i {e_{i\alpha}} = (\sum_i {\bf M}_i)
{u_{\alpha}},
\label{eq9}
\end{equation}
which ensures that the fluid minimizes its free energy at equilibrium.

Conditions (\ref{eq6})--(\ref{eq9})
can be satisfied as is usual in lattice Boltzmann
schemes by writing the equilibrium distribution functions and forcing
terms as polynomial expansions in the velocity\cite{CD98}
\begin{eqnarray}
f_i^{eq}&=&A_s + B_s u_\alpha e_{i\alpha}+C_s u^2+D_s u_\alpha
u_\beta
e_{i\alpha}e_{i\beta}+E_{s\alpha\beta}e_{i\alpha}e_{i\beta},\nonumber \\
{\bf G}_i^{eq}&=&{\bf J}_s + {\bf K}_s u_\alpha e_{i\alpha}+{\bf L}_s
u^2+{\bf N}_s u_\alpha
u_\beta e_{i\alpha}e_{i\beta},\nonumber \\ 
p_i&=&T_s \partial_\beta \tau_{\alpha\beta} e_{i\alpha},\nonumber \\ 
{\bf M}_i&=&{\bf R}_s+{\bf S}_s u_\alpha e_{i\alpha},
\label{expansion_in_u}
\end{eqnarray}
where $s={\vec e}_i\;^2 \in \{0,1,2\}$ identifies separate coefficients
for different absolute values of the velocities.  A suitable choice is
\begin{eqnarray}
&&A_2=-(\sigma_{xx}+\sigma_{yy})/16,\qquad A_1=2 A_2,\qquad A_0=\rho-12 A_2,\nonumber \\
&&B_2=\rho/12,\qquad B_1=4 B_2,\nonumber \\ &&C_2=-\rho/16,\qquad C_1=-\rho/8,
\qquad C_0=-3 \rho/4,\nonumber \\ &&D_2=\rho/8, \qquad D_1=\rho/2\nonumber \\
&&E_{2xx}=(\sigma_{yy}-\sigma_{xx})/16,\qquad E_{2yy}=-E_{2xx}, \qquad
E_{2xy}=E_{2yx}=-\sigma_{xy}/8,\nonumber \\ &&E_{1xx}=4 E_{2xx},\qquad E_{1yy}=4
E_{2yy},\nonumber \\ &&{\bf J}_0={\bf Q},\nonumber \\ 
&&{\bf K}_2={\bf Q}/12,\qquad {\bf K}_1=4 {\bf K}_2,\nonumber \\
&&{\bf L}_2=-{\bf Q}/16,\qquad {\bf L}_1=-{\bf Q}/8, \qquad {\bf
L}_0=-3 {\bf Q}/4,\nonumber \\ 
&&{\bf N}_2={\bf Q}/8, \qquad
{\bf N}_1={\bf Q}/2\nonumber \\ 
&&T_2=1/12, \qquad T_1=4 T_2,\nonumber \\ 
&&{\bf R}_2=\widehat{\bf H}/9, \qquad
{\bf R}_1={\bf R}_0={\bf R}_2\nonumber \\ &&{\bf S}_2=\widehat{\bf H}/12, \qquad {\bf S}_1=4 {\bf S}_2,
\label{eq14}
\end{eqnarray}
where any coefficients not listed are zero.

\section{H-V Phase diagram}

\begin{figure}
\vskip -1.5true cm
\begin{picture}(150,50)(-50,60)
\put(0,0){\vector(1,0){50}}
\put(0,0){\vector(0,-1){50}}
\put(50,-10){$y$}
\put(-10,-60){$x$}
\end{picture}
\centerline{\epsfxsize=2.5in \epsffile{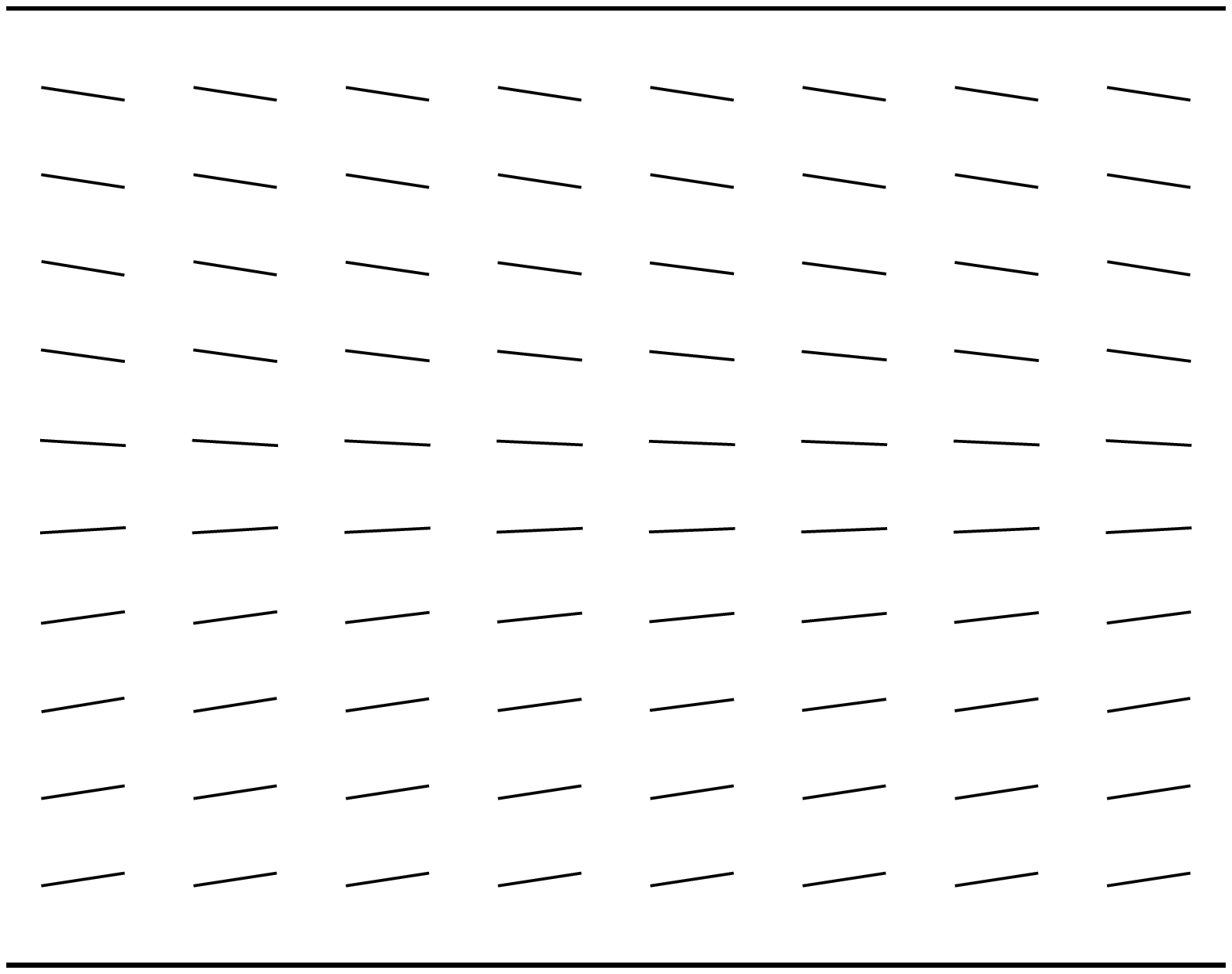}}
\vskip -0.5true cm
\centerline{(a)}
\centerline{\epsfxsize=2.5in \epsffile{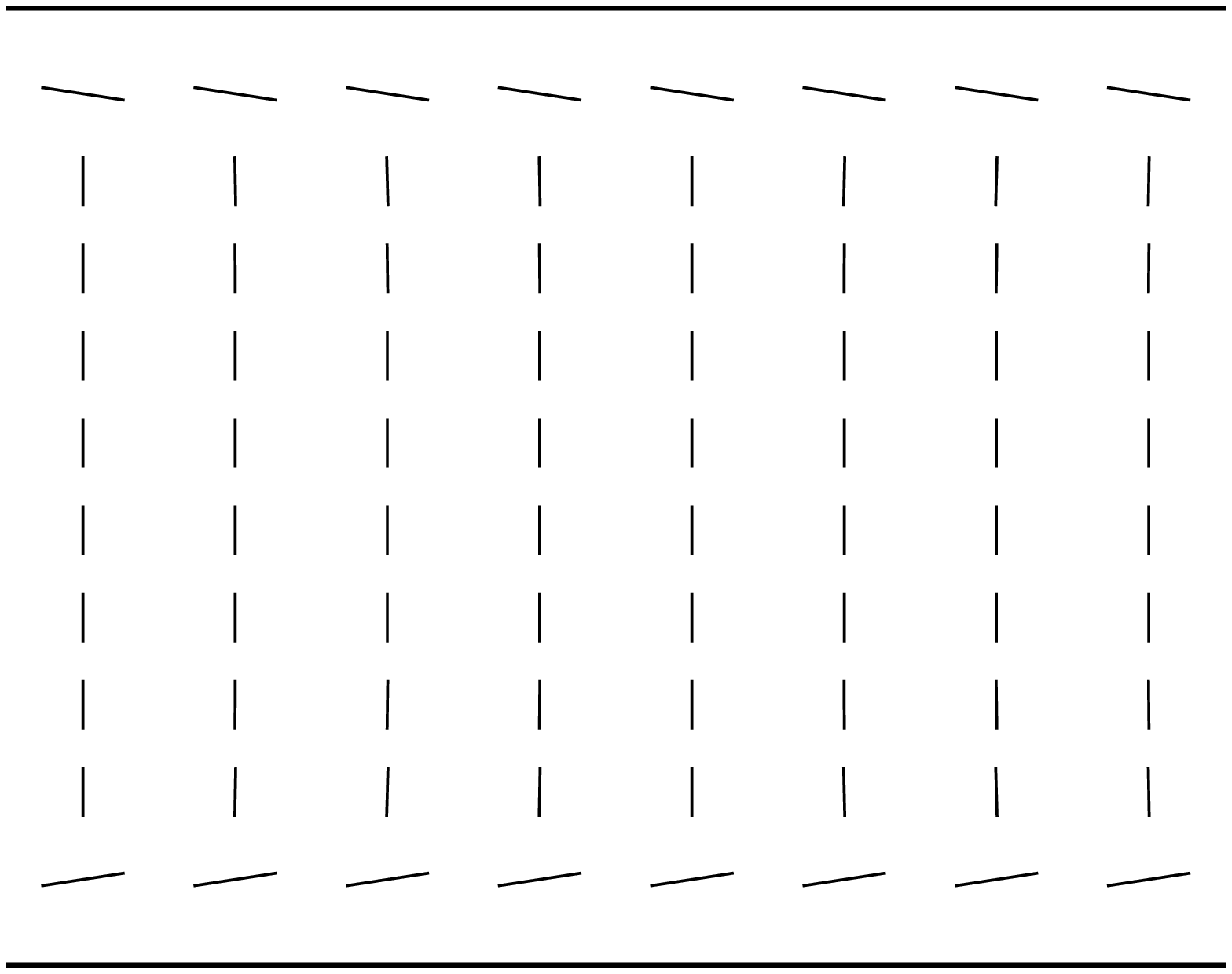}}
\vskip -0.5true cm
\centerline{(b)}
\centerline{\epsfxsize=2.5in \epsffile{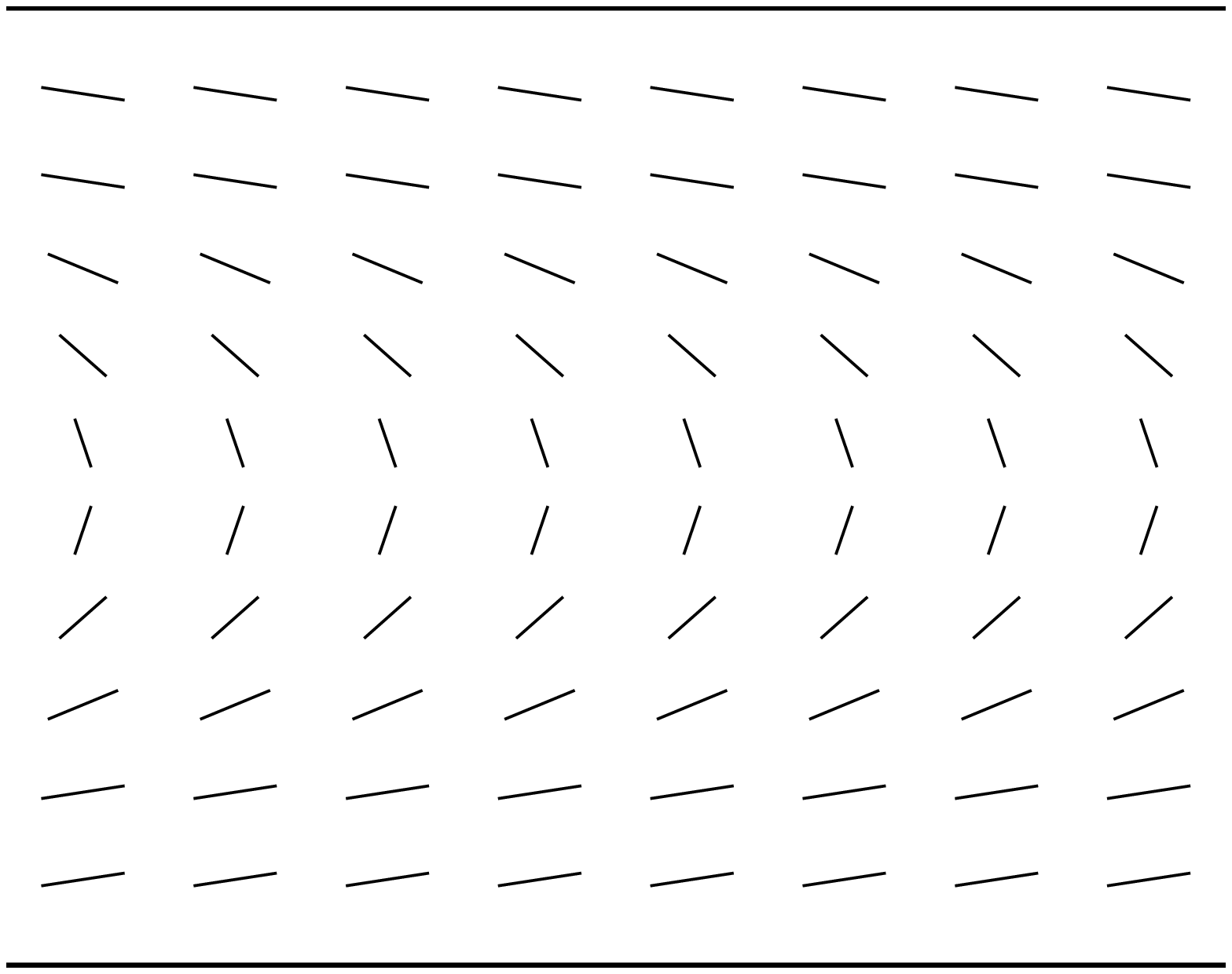}}
\vskip -0.5true cm
\centerline{(c)}
\vskip 1.0true cm
\caption{}
\label{fig_surf2}
\end{figure}

We now use the lattice Boltzmann algorithm described in Sections 2 and 3 to
investigate a simple liquid crystal device. The device consists of a
nematic liquid crystal confined between two planes a distance $L_x$
apart.  At the surface ${\bf Q}^0$ is chosen to give a surface tilt to
the director of $+\theta_p$ at $x=0$ and  $-\theta_p$ at $x=L_x$.  
At zero applied voltage these conditions result in a global minimum
free energy state with a splayed director configuration, or H state as
shown in Figure~\ref {fig_surf2}(a).  At high voltages, typically on
order of $~6 V$, the H state is no longer the global minimum, and a
bend configuration (V state) is obtained like the one shown in 
Figure~\ref {fig_surf2}(b).  At intermediate voltages, the V state is
more relaxed, like the one shown in Figure~\ref {fig_surf2}(c).  

The V state may remain for some time even at zero voltage as it is metastable.
As the H and V states are topologically distinct, the transition from one
to the other requires nucleation and the generation of defects.
This process is illustrated in Figure~\ref{fig_twodefects}.

The simulations were performed on a $48 \times 48$ grid.
We used $L_1 = 0.0440$, $L_2 = 0.0445$, $L_3 = 0.0606$, 
$\epsilon_a=41.4$, $\epsilon_m=9.8$,
$A_0=1.0$ and $\gamma=3.0$.  By a choice of appropriate length, time and 
pressure scales these correspond to physical values 
values $L_1 = 17.4$ pN, $L_2 = 17.6$ pN,
$L_3 = 24.2$ pN, 
$L_x=L_y=3$ $\mu m$ which is consistent with the
nematic liquid crystal E7 (Merck Ltd. UK) at $25^\circ C$.
Periodic boundary
conditions were used in the $y$ direction and bounce-back in the $x$
direction. 

A phase diagram for the H and V states in the tilt angle/voltage plane
is shown in Fig. \ref{phasediag}. The system was started in a configuration
like that shown Figure~\ref{fig_twodefects}. For low fields the H domain has 
a lower free energy density, thus it grows (crosses). 
For high enough fields the H 
domain shrinks (circles). 
The boundary is defined as the voltage at which the two domains
have the same free energy. This diagram compares well with those measured
experimentally for a system using E7 \cite{AT00}.

\section{Domain growth}

Consider the dynamics of the domain growth in the system.
Rather than fix parameters to a specific material, we wish to explore
how the dynamics are affected by different parameters.  In particular,
we are interested in how hydrodynamics affects the speed of the
domain growth.  This is motivated by the observation in
Ref. \cite{AT00} that the domain growth was anisotropic and the speculation
that this may be due to hydrodynamics.  To isolate effects
due to different elastic constants, for these simulations we set
$L_2=L_3=0.0$, the so-called one-elastic constant approximation. 

The initial configuration, depicted in Fig. \ref{fig_twodefects}(a), 
is a horizontal (i.e., along the $y$-direction) domain
in an otherwise vertically aligned state. This models a time shortly
after the electric field has been switched off when small but
macroscopic domains have formed in the device. 

Once the simulation commences the director configuration relaxes
rapidly to that shown in Fig. \ref{fig_twodefects}(b). 
The horizontal and vertical domains
are both tilted slightly by the elastic coupling to the surface
spins. Then defects are formed at the domain boundaries, which merge
to form single defects with strengths $\pm 1/2$ at the centre of each
domain wall respectively. Once the two defects have formed 
the vertical domain begins to grow
and the defects move in opposite directions. Our aim is to investigate the
effect of back-flow and surface director tilt on this
motion. Most previous work on the dynamics of topological defects
has ignored the back-flow \cite{D96,PR92,RK91,JM96}.

We first investigate the effect of the surface tilt $\theta_p$ on the defect
speed.  The horizontal domain grows because it decreases 
the free energy of the device. It can be shown \cite{AT00} that
the difference in free energy between the horizontal and vertical domains 
is proportional to $45^\circ-\theta_p$, where $\theta_p$ is the angle
of the surface tilt to the $y$-axis. Thus as $\theta_p$ increases, 
the free energy difference decreases,
and the defects move more slowly. At $\theta_p=45^\circ$ 
the two domains have the same free energy and the defects
stop moving. For $\theta_p>45^\circ$ the horizontal domain begins to shrink.

A particular advantage of the simulation is that the
back-flow can easily be switched off by setting 
$\sigma_{\alpha \beta}=-P_0 \delta_{\alpha \beta}$ and 
$\tau_{\alpha \beta}$ to zero in (\ref{expansion_in_u}) and (\ref{eq14}). 
Thus the effect of the back-flow can be unambiguously identified.
Since no flow is imposed, this
leads to a zero velocity field throughout the whole simulation. 
The dynamical equation for the system in this case can be obtained
from (\ref{Qevolution}) by setting ${\vec u}$ to zero.
It corresponds to a simpler model which does not include hydrodynamics.

Consider first the diamonds in Fig. \ref{fig_defectspeed_tilt}. 
This corresponds to the case with back-flow switched off. For this case
both defects move at the same speed (but in opposite directions). This
can be explained by considering a local co-ordinate transformation  
$x \rightarrow -x$ which is equivalent to changing the sign of the
off-diagonal elements in ${\bf Q}$. Assuming  ${\vec u}=0$, 
the dynamical equation
(\ref{Qevolution}) is invariant under this transformation whereas the
two defects in Fig. \ref{fig_twodefects}(b) transform into each
other. 

The triangles and circles in Fig. \ref{fig_defectspeed_tilt} show the
velocity of the defects when back-flow is included in the model. 
Now the two defects move with different speeds since the $xy$ elements
of the stress tensors (\ref{BEstress}) and (\ref{as}) are not
invariant under the $x \rightarrow -x$ transformation. Thus the stress
fields are different for the two defects.  Due to this stress field
two different flow fields are formed around the defects (see
Fig. \ref{fig_velocityfield}). The flow is
stronger around the $s= + \frac {1} {2}$ defect and it is
substantially accelerated. The speed of the $s= - \frac {1} {2}$ defect
is affected much less by the flow.

An experimental setup similar to Fig. \ref{fig_twodefects}  
was considered in \cite{AT00} where the growth of a circular twist
domain in a horizontal environment under the influence of an external
field was studied for different surface tilt values. It was found that
the domain growth was anisotropic. Although this problem is
three-dimensional, a vertical cross section through the growing domain
gives a geometry similar to that considered here. It seems extremely
plausible, that the essential physics is captured by our model with
the anisotropy in growth resulting from the difference in the flow
fields corresponding to different defect topologies at the defect boundary.
Quantitative mappings between the experiment and the simulation
results must be treated with caution because of the different
dimensionality. However, in both the simulation and the experiment
the difference between the wall 
speeds was about $40-50\%$.

The simulations were performed on a $700 \times 48$ grid.
We used $L_1 = 1.0$, $L_2 = L_3 = 0$, $A_0=5.0$, $\gamma=3.0$,
$\Gamma=0.44$, $\xi=0.52$ and $\tau_f=1.0$.
The larger $L_1$ corresponds to increasing the resolution
of the lattice in order to perform accurate defect
velocity measurements.

\section{Summary and Discussion}

In this paper we have described a lattice Boltzmann
algorithm to simulate the dynamics of a non-Newtonian fluid, 
liquid crystals.
 
In particular we have considered the the presence
of an external electric field and the possibility of 
a general Frank free energy with three elastic constants.  In the 
continuum limit we recover the Beris-Edwards formulation within which the 
liquid crystal equations of motion are written in terms of a
tensor order parameter \cite{BE94_BE90}. The equations are 
applicable to the isotropic, uniaxial nematic, and biaxial 
nematic phases.  Working within the framework of a variable tensor 
order parameter it is possible to simulate variations
in the magnitude of order and hence the dynamics of 
topological defects and non-equilibrium phase transitions between 
different flow regimes.

The algorithm was used to explore states in nematic liquid crystal
devices. We find that switching between topologically distinct states
is, as expected, strongly inhibited by a free energy barrier between
the initial and the final states.  
In this case the state can only change once a domain of the final state 
has been nucleated. Defects form at the moving domain walls
and hence this geometry allows an investigation of defect hydrodynamics. 
We find that a $s=+\frac {1} {2}$ defect is substantially speeded up
by back-flow effects.
This is relevant to device physics where the speed
of switching is an important design variable.

There are many directions for further research opened up by the rich
physics inherent in liquid crystal hydrodynamics and the generality of
the Beris-Edwards equations. For example, another possible switching
pathway would be via the escape of the director into a third dimension.
The addition of flexoelectric terms to the equations of motion will allow problems relevant to bistable liquid crystal displays to be addressed. 
Even cases where the switching is the consequence of back-flow \cite{DN97}
can be modeled.
Numerical investigations are vital to address these problems 
because of the complexity of the equations of motion.

\eject

FIGURE CAPTIONS

Fig. 1 Alignment of molecules for a tilt angle $+\theta_p$ on the top
surface and $-\theta_p$ on the bottom surface. 
Director configuration when $\theta_p < 45^\circ$ and (a) the field is
switched off and the system has had time to relax to its global
minimum; (b) the field is switched on the a fairly high voltage 
$ \sim 6 V$;
and (c) the field is at a low voltage $ \sim 2 V$ or lower.  The system
may remain in the metastable state (c) for some time even at zero voltage. 

Fig 2. (a) Initial director configuration used to study domain growth.
After the
vertical electric field is switched off, a horizontal domain is created
in order to nucleate the growth process.
(b) Shortly thereafter two defects are formed at 
the boundary of the horizontal and vertical domains. The left (right)
defect has a topological strength $s= -\frac {1} {2}$ ($s= +\frac {1} {2}$). 
Both the horizontal and the vertical domains are
slightly distorted due to the defects and the surface tilt.

Fig. 3. Phase diagram for the H and V states as a function of tilt
  angle and voltage. At low fields (crosses) the H domain has a 
lower free energy 
density, thus it grows. If the field strong enough (circles), 
the H domain shrinks. On the line the two domains have equal free
energy density.

Fig. 4. The velocity of the two defects as the function of surface tilt if back-flow is ignored (diamonds)
and included.  Notice that if back-flow is not included  then the two defects move with the same speed. 
Hydrodynamics accelerates the $s= +\frac {1} {2}$ defect (triangles) substantially, 
while it affects less the $s= -\frac {1} {2}$ defect (circles).

\begin{figure}
\centerline{\epsfxsize=5.in \epsffile{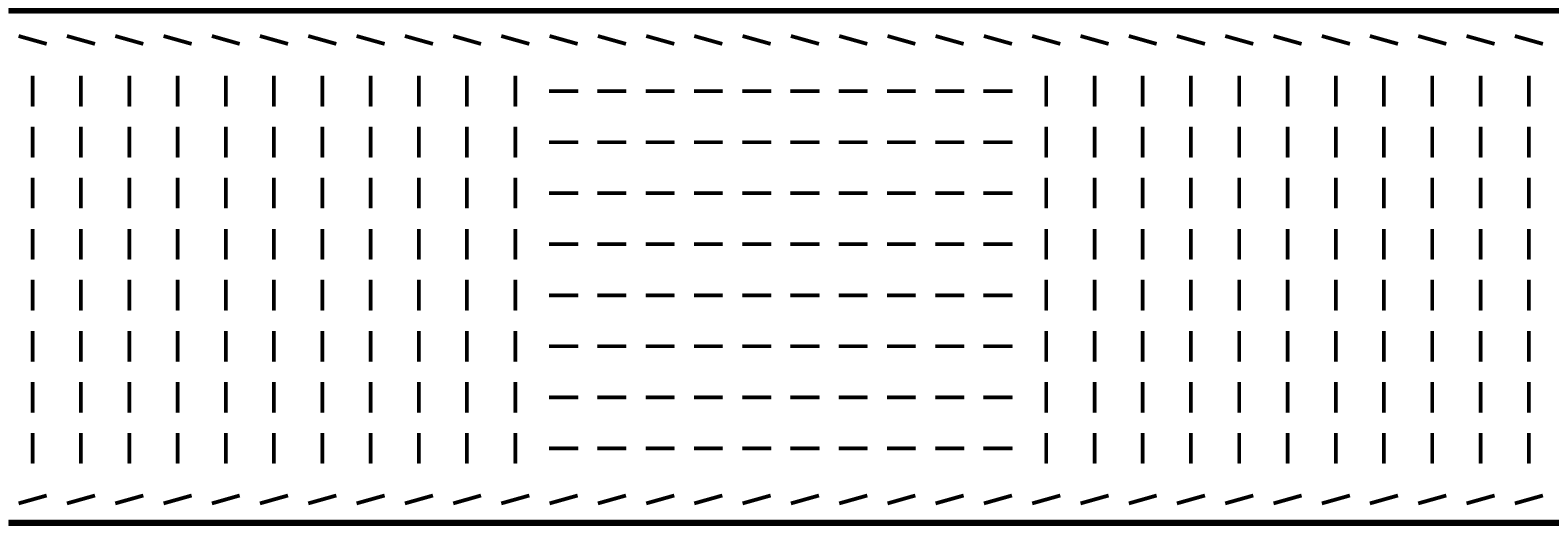}}
\centerline{(a)}
\vskip 0.5true cm
\centerline{\epsfxsize=5.in \epsffile{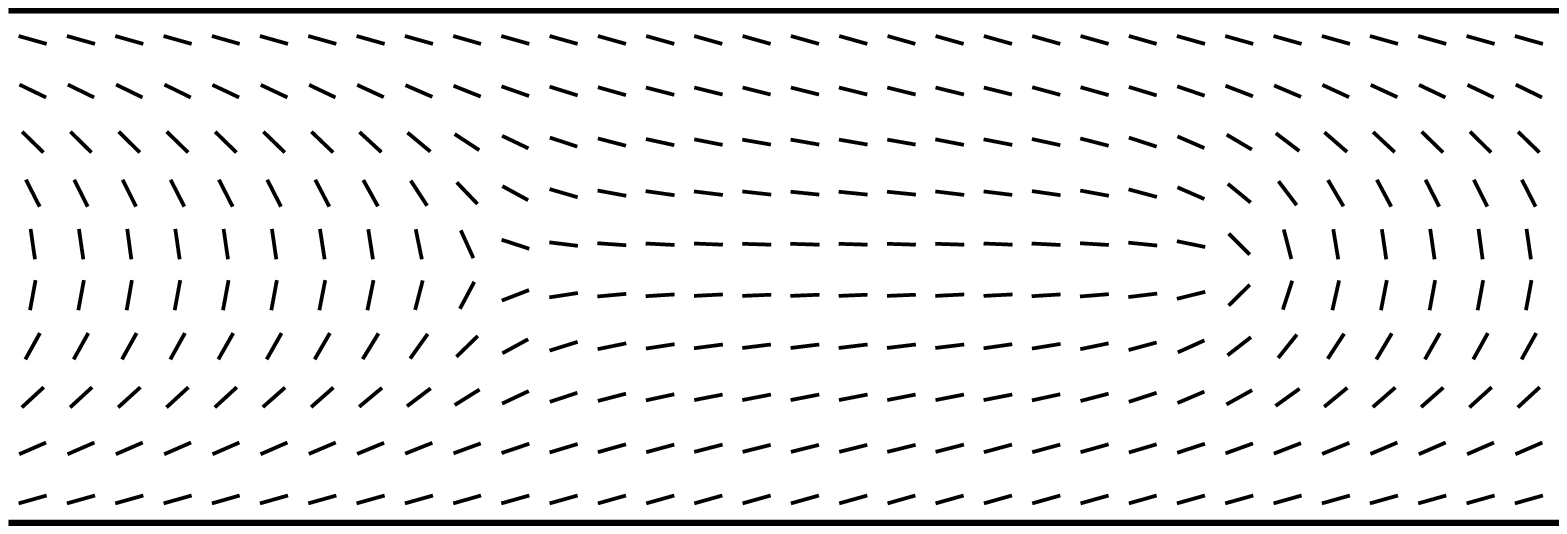}}
\centerline{(b)}
\vskip 1.0true cm
\caption{}
\label{fig_twodefects}
\end{figure}

\begin{figure}
\centerline{\epsfxsize=5.in \epsffile{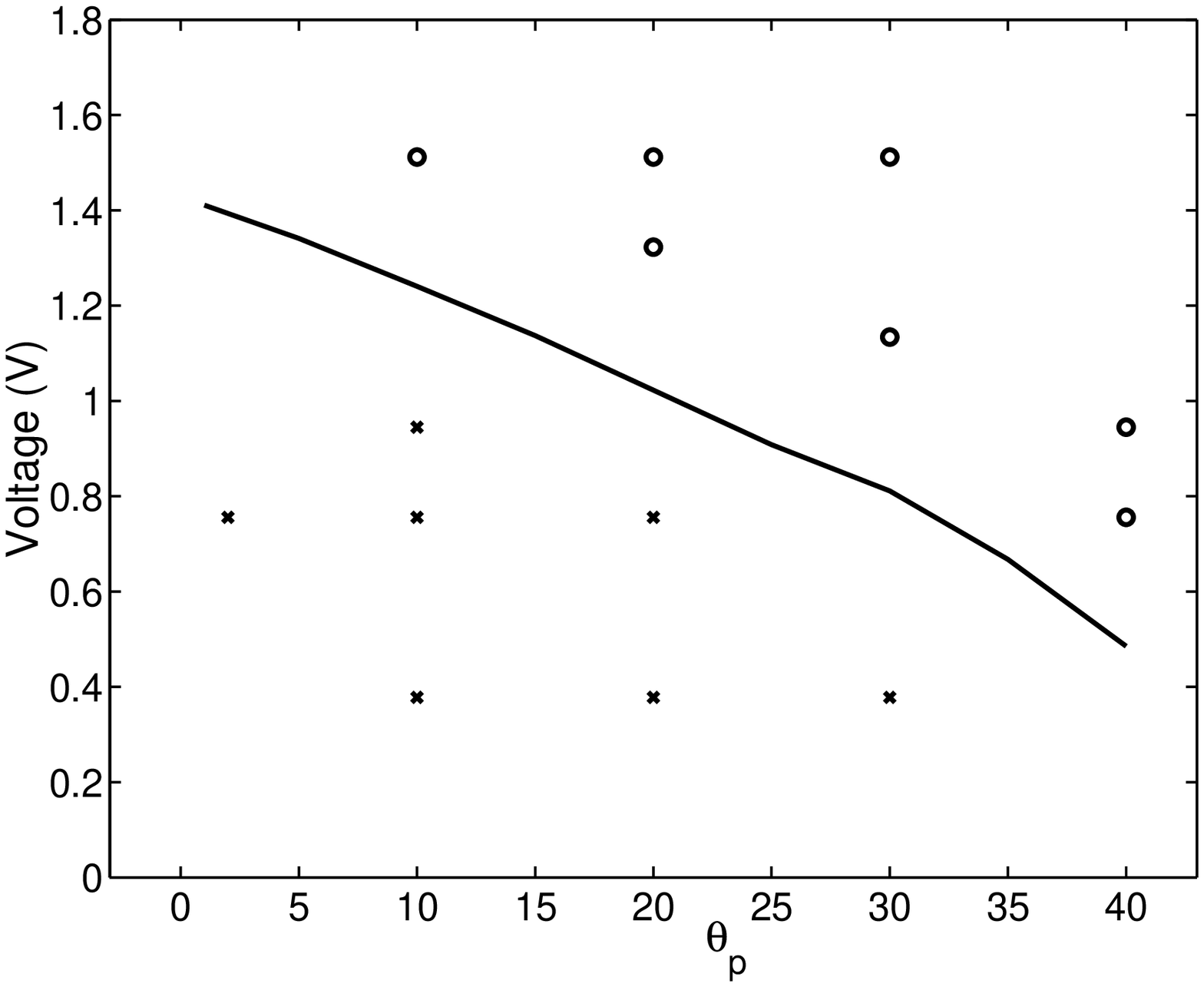} }
\vskip 0.8true cm
\caption{}

\label{phasediag}
\end{figure}

\begin{figure}
\centerline{\epsfxsize=5.in \epsffile{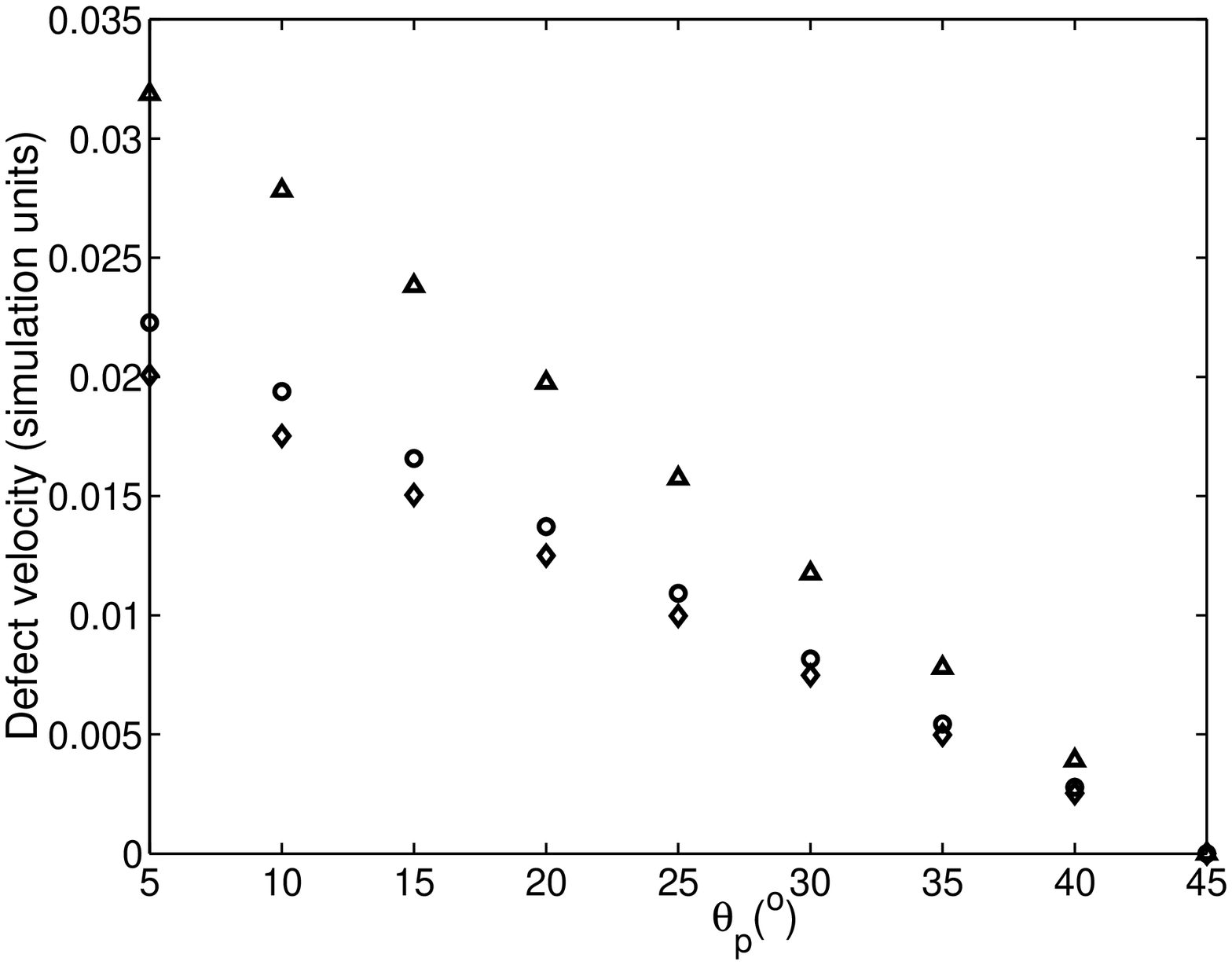} }
\vskip 0.8true cm
\caption{}
\label{fig_defectspeed_tilt}
\end{figure}

\begin{figure}
\centerline{\epsfxsize=6.in \epsffile{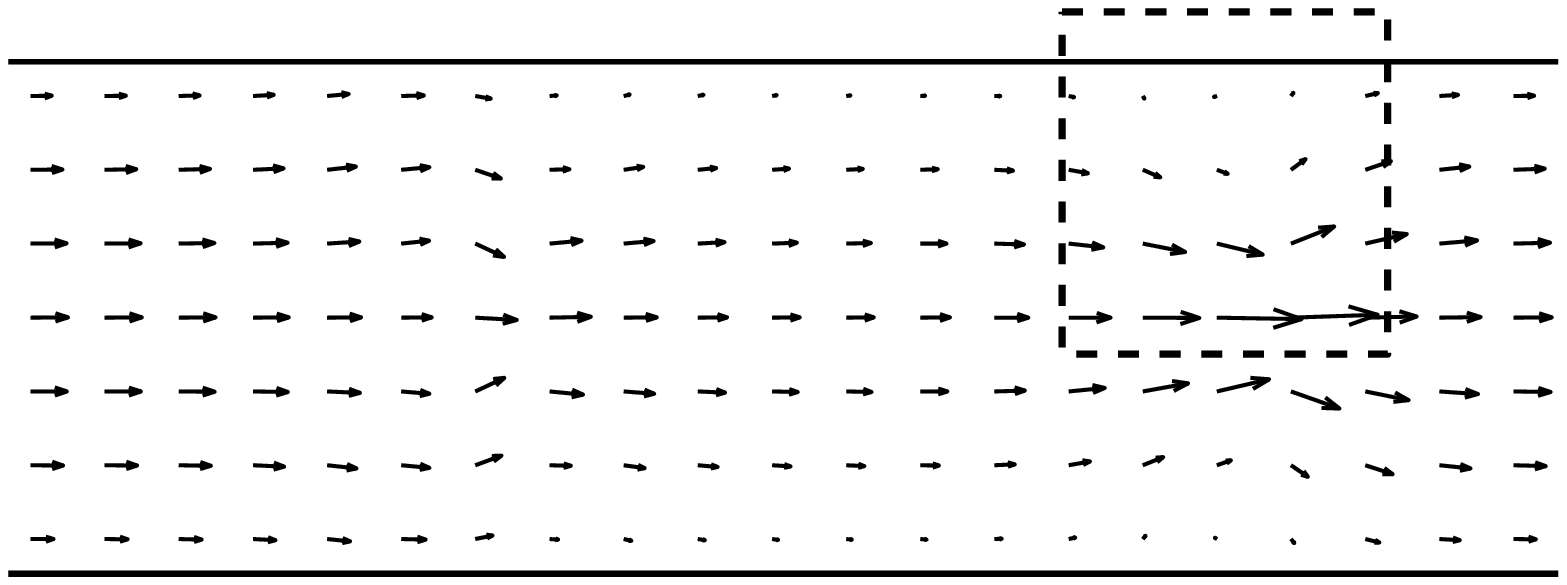} }
\vskip -0.2true cm
\centerline{(a)}
\vskip  0.5true cm
\centerline{\epsfxsize=3.in \epsffile{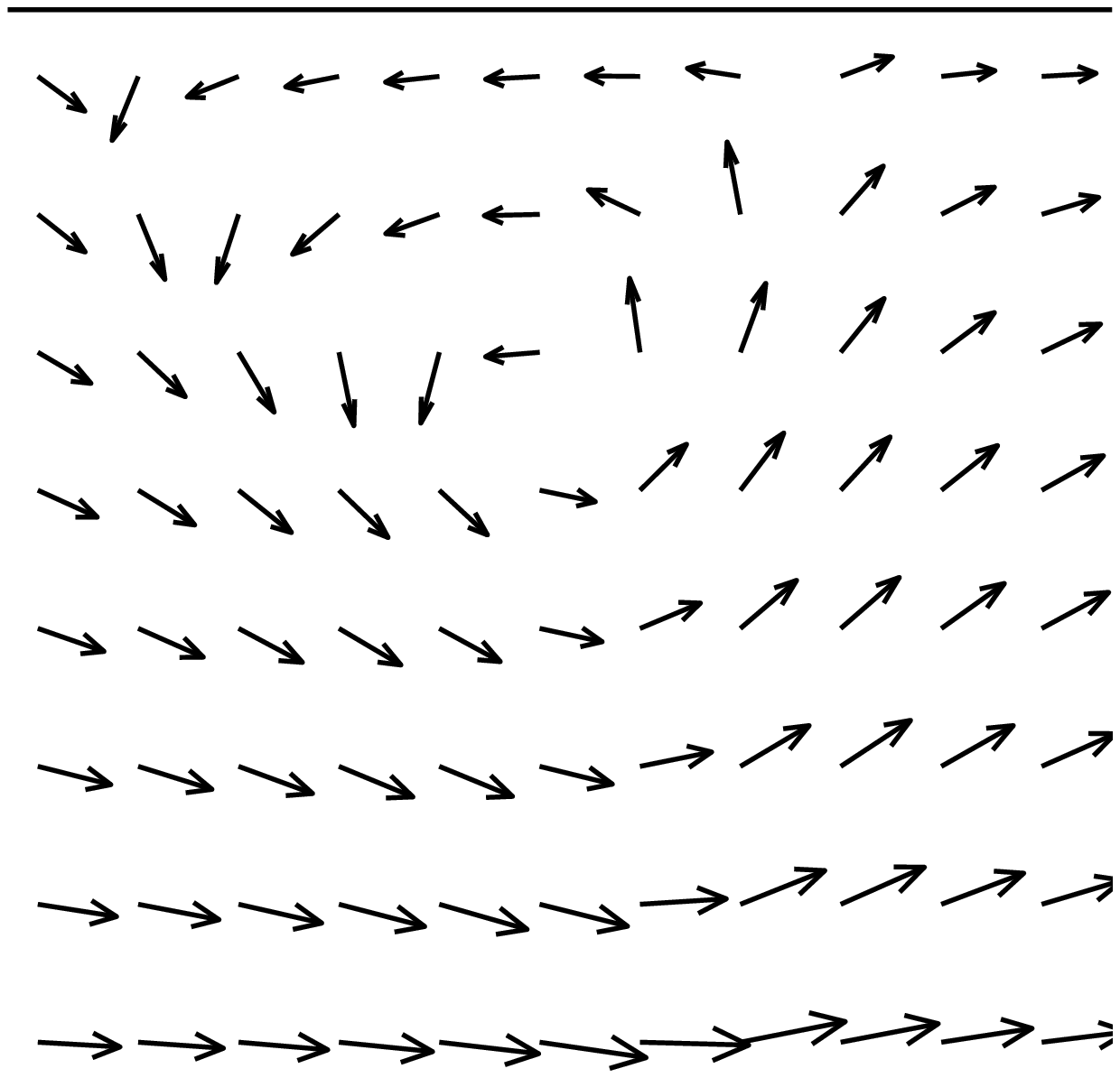} }
\vskip -0.2true cm
\centerline{(b)}
\vskip 0.2true cm
\caption{}

\label{fig_velocityfield}
\end{figure}

Fig. 5. (a) Velocity field corresponding to the director configuration shown 
in Fig. \ref {fig_twodefects}(b).  
Notice that there is a strong vortex structure around 
the $s= +\frac {1} {2}$ defect
and at the defect core it points in the direction of defect movement. 
The flow at the core of the other defect is much weaker. 
(b) Flow field corresponding to the dashed box. For better visibility the 
length of the vectors is changed disproportionally.

\eject

\eject 

\eject

\eject

\end{document}